\documentclass[aps, showpacs, draft, eqsecnum, superscriptaddress]{revtex4}

\usepackage{epsf}  
\usepackage[final]{graphicx}  
\usepackage{amsmath,amssymb} 

\begin{document}

\title{Toward standard testbeds for numerical relativity}
%

\author{Miguel Alcubierre}
\affiliation{Instituto de Ciencias Nucleares, Universidad Nacional Aut\'onoma
de M\'exico, Apartado Postal 70-543, M\'exico Distrito Federal 04510,
M\'exico.}

\author{Gabrielle Allen}
\affiliation{
        Max-Planck-Institut f\"ur Gravitationsphysik,
        Albert-Einstein-Institut,
        14476 Golm, Germany}

\author{Carles Bona}
\affiliation{Departament de Fisica, Universitat de les
Illes Balears,
Ctra de Valldemossa km 7.5, 07122 Palma de Mallorca,
Spain}

\author{David Fiske}
\affiliation{Department of Physics,
        University of Maryland,
        College Park, MD 20742-4111, USA}

\author{Tom Goodale}
\affiliation{
        Max-Planck-Institut f\"ur Gravitationsphysik, 
        Albert-Einstein-Institut,
        14476 Golm, Germany}

\author{F.~Siddhartha Guzm\'an}
\affiliation{
        Max-Planck-Institut f\"ur Gravitationsphysik, 
        Albert-Einstein-Institut,
        14476 Golm, Germany}

\author{Ian Hawke}
\affiliation{
        Max-Planck-Institut f\"ur Gravitationsphysik, 
        Albert-Einstein-Institut,
        14476 Golm, Germany}

\author{Scott H. Hawley}
\affiliation{
        Center for Relativity,
        University of Texas at Austin,
        Austin, Texas 78712 USA}

\author{Sascha Husa}
\affiliation{
        Max-Planck-Institut f\"ur Gravitationsphysik, 
        Albert-Einstein-Institut,
        14476 Golm, Germany}

\author{Michael Koppitz}
\affiliation{
        Max-Planck-Institut f\"ur Gravitationsphysik, 
        Albert-Einstein-Institut,
        14476 Golm, Germany}

\author{Christiane Lechner}
\affiliation{
        Max-Planck-Institut f\"ur Gravitationsphysik, 
        Albert-Einstein-Institut,
        14476 Golm, Germany}

\author{Denis Pollney}
\affiliation{
        Max-Planck-Institut f\"ur Gravitationsphysik, 
        Albert-Einstein-Institut,
        14476 Golm, Germany}

\author{David Rideout}
\affiliation{
        Max-Planck-Institut f\"ur Gravitationsphysik, 
        Albert-Einstein-Institut,
        14476 Golm, Germany}

\author{Marcelo Salgado}
\affiliation{Instituto de Ciencias Nucleares, Universidad Nacional Aut\'onoma
de M\'exico, Apartado Postal 70-543, M\'exico Distrito Federal 04510,
M\'exico.}

\author{Erik Schnetter}
\affiliation{
        Institut f\"ur Astronomie und Astrophysik, 
        Universit\"at T\"ubingen, 72076 T\"ubingen, Germany}

\author{Edward Seidel}
\affiliation{
        Max-Planck-Institut f\"ur Gravitationsphysik, 
        Albert-Einstein-Institut,
        14476 Golm, Germany}

\author{Hisa-aki Shinkai}
\affiliation{
        Computational Science Division, 
        Institute of Physical \& Chemical Research (RIKEN), 
        Hirosawa 2-1, Wako, Saitama, 351-0198 Japan}

\author{Deirdre Shoemaker}
\affiliation{
        Center for Radiophysics and Space Research,
        Cornell University, Ithaca, NY 14853, USA}

\author{B\'ela Szil\'agyi}
\affiliation{
        Department of Physics and Astronomy,
        University of Pittsburgh, Pittsburgh, PA 15260, USA}

\author{Ryoji Takahashi}
\affiliation{Theoretical Astrophysics Center, 
        Juliane Maries Vej 30, 
        2100 Copenhagen, Denmark }

\author{Jeff Winicour}
\affiliation{
        Department of Physics and Astronomy,
        University of Pittsburgh, Pittsburgh, PA 15260, USA}
\affiliation{
        Max-Planck-Institut f\"ur Gravitationsphysik, 
        Albert-Einstein-Institut,
        14476 Golm, Germany}

\collaboration{Mexico Numerical Relativity Workshop 2002 Participants} 
\noaffiliation

\date{\today}

\begin{abstract}
In recent years, many different numerical evolution schemes for Einstein's
equations have been proposed to address stability and accuracy problems that
have plagued the numerical relativity community for decades. Some of these
approaches have been tested on different spacetimes, and conclusions have been
drawn based on these tests. However, differences in results originate from many
sources, including not only formulations of the equations, but also gauges,
boundary conditions, numerical methods, and so on. We propose to build up a
suite of standardized testbeds for comparing approaches to the numerical
evolution of Einstein's equations that are designed to both probe their
strengths and weaknesses and to separate out different effects, and their
causes, seen in the results. We discuss general design principles of suitable
testbeds, and we present an initial round of simple tests with periodic
boundary conditions. This is a pivotal first step toward building a suite of
testbeds to serve the numerical relativists and researchers from related
fields who wish to assess the capabilities of numerical relativity
codes. We present some examples of how these tests can be quite effective in
revealing various limitations of different approaches, and illustrating their
differences. The tests are presently limited to vacuum spacetimes, can be run on
modest computational resources, and can be used with many different approaches
used in the relativity community.
\end{abstract}

\maketitle

\section{Introduction}

The inspiral of a relativistic binary has played the role of a standard candle
for the first signal to be detected by the gravitational wave observatories
which are now approaching operational readiness. For many years now, this has
spurred activity to simulate the inspiral and merger of binary black holes
using  fully
three-dimensional general relativistic evolution codes. Several groups
across the world are dedicated to this endeavor but it still lies beyond
present capability. The reasons for the difficulty of the binary black hole
problem reflect the complexity of the underlying physics: the computational
domain has a geometry whose metric is highly dynamic on vastly different time
scales in the inner and outer regions, and it has a topology which is subject
to change in order to avoid singularities. Even in the absence of black holes,
there is no consensus on the best analytic and geometric formalism for dealing
with the nonlinear nature of the gauge freedom and constraints of general
relativity, and a numerical treatment greatly compounds the possibilities. The
problem poses an enormous scientific challenge. See recent reviews
such as~\cite{Baumgarte:2002jm,Shinkai02a,Lehner01a}. 

Substantial progress on small pieces of the problem has been made by individual
groups but that progress is difficult to assess on a community level for the
purpose of sharing ideas. A newcomer to the field would have to rely completely
on anecdotal evidence in deciding how to develop a code. An observer or
phenomenologist cannot judge the status of a code by wading through pages of
Fortran and needs to view some standard tests to assess its reliability.
Results are often published by various groups that report improvements in
stability or accuracy in evolutions.; but formulations, gauges,  numerical
methods, boundary conditions and initial data can be inextricably mixed, and it
is often difficult to sort out which ingredients in an improved treatment are
crucial and against what standard the  improvement is measured.  Different
groups invariably use different criteria for nearly all these issues.  This
paper is a first step toward remedying this situation by the establishment of
standard testbeds that will help the numerical relativity community present and
share results in an objective and useful way that will further scientific
progress.

Apart from the practical aspects of providing a collection of standard tests as
a community resource, the formulation of an appropriate test suite presents a
significant scientific challenge. Comparing codes based upon different sets of
variables with different sets of constraint equations is a nontrivial task. It
is not straightforward to draw conclusions which would extend beyond the
particular simulation for which the comparison was made. It is important to
remain aware that evolution systems are a unity of evolution equations,
boundary conditions and gauge. In order to paint a picture which gives
heuristic insight into a broad  panorama of ``successes'' and ``failures'', it
is essential to carefully design a set of tests involving different spacetimes
and gauges. The current paper is a pivotal first step in this direction.
Although it does not present a complete test suite, it provides some foundation
for building one. We envisage the tests proposed here as the first round in a
series of increasingly more complex  tests. 

In order to facilitate comparison of results, we will specify all tests
as explicitly as possible. In addition to describing initial data and
gauge, we will also specify a minimal set of output quantities,
the setup of numerical grids, the choice of resolutions for convergence
testing and even numerical methods.
Although our choices aim at broad applicability, it is clear that they will
not be optimal for many formulations. We foresee
the usefulness of additional results, obtained with improved numerical 
methods or other modifications in the setup, which may promote further insight
and clarification.
We stress however, that results produced in a simple and identical numerical
setup form the essential basis for a numerical comparison of different 
formulations of the equations. We present sample test results which illustrate
and motivate such specifications.
In this initial round of proposed tests, we also keep the spatial resolution
rather low so as not to strain the computational resources of any group.

We believe that this project
introduces a natural framework for documenting algorithms that ``do not work'',
which so far has been sorely lacking but is important for avoiding redundant
experimentation. While failure in one particular case might be easy to detect,
the general task of exploring and comparing different codes over a wide set of
situations is complex. 

Already nontrivial is the selection of criteria for comparison. Some standard
criteria are: 
\begin{enumerate} 
\item {\em Stability}: Exponential growth
(unless on a time scale significantly larger than the relevant physical
timescale) is usually not tolerable in numerical simulations and should not
occur if not inherent in the analytic problem. 
\item {\em Accuracy}: Apart from
resolution and numerical methods (which may be restricted by available
computational resources), accuracy also depends on the analytic formulation.
\item {\em Robustness}: From a practical point of view, one is interested in a
variety of physical situations. A robust code should be able to perform well in
a wide class of spacetimes, gauges etc. 
\item {\em Efficiency}: For 3D
numerical relativity, time and memory constraints are severe. Thus
computational efficiency is a serious issue. 
\item {\em Degree of mathematical
understanding}: Optimally one would like to mathematically demonstrate certain
features of evolution systems, such as well-posedness or von Neumann stability.
\end{enumerate}

In this work we focus on the issue of stability, and to some degree on
accuracy. By devising a broad set of tests, we also shed some light on the
question of robustness. At this stage, we do {\em not} investigate efficiency,
i.e.\  we do not consider our tests as benchmarks of computational performance
regarding speed or memory requirements.  Also, we adopt a practical, empirical
point of view of what works and what does not, regardless of whether a
mathematical theory is available. We hope, of  course, that some of our results
might stimulate theoretical progress. Indeed we hope that our work is not only
useful for numerical relativists but also for a more general audience of
relativists who are interested in the current state of numerical simulations.
Furthermore, this work could be useful in facilitating  communication between
numerical relativists and the broad community of computational mathematicians
and physicists.

The ideas presented here are an outgrowth of the ``Apples with Apples''
numerical relativity workshop held at UNAM in Mexico City during May 2002. The
objective of the workshop was to formulate some basic tests that could be
carried out by any group with a 3-dimensional code. What we report here are
details of a selection of the first tests that were proposed and the principles
that went into their design. The test results will be published at a later
time. We encourage all groups, whether they attended the workshop or not, to
contribute test results and share authorship in this second publication. While
some tests may not be easily implemented in all formalisms, it is important for
the calibration and support of progress in the field that the various groups
submit as many test results as practical and conform as closely as possible to
the test specifications. A web site www.ApplesWithApples.org has been set up to
coordinate efforts and display results. Specific information about submitting
tests results, and accessing results from other groups, can be obtained there.
A continued series of workshops is being planned with the goal of developing
increasingly more demanding tests leading up to black hole spacetimes with grid
boundaries. The web site provides a forum for proposing new tests and
coordinating plans for future workshops.

The round of preliminary tests presented here is of a simple nature designed to
facilitate broad community participation.  All tests use periodic boundary
conditions, which is equivalent to evolution on the 3-torus in the absence of
boundaries. Because the treatment of grid boundaries is one of the most serious
open problems in numerical relativity, it is extremely useful to look in detail
at this case which is not obscured by boundary effects. Only the Gowdy wave
test is based on genuinely nonlinear data. The other tests involve weak and
moderately nonlinear fields. Successful performance of an evolution code at
this level is not obscured by either strong field effects or boundary effects
and is indeed imperative for the successful simulation of black holes.  The
tests can be readily performed by any group with the capability of conducting
fully 3-dimensional simulations. The value of this project depends critically
on ease of implementation and flexibility of expansion to future tests.

The main purpose of the tests is to provide a framework for assessing the
accuracy and long term stability of simulations based upon the wide variety of
formulations and numerical approaches being pursued. All present codes are in a
state of flux. The community needs clear information about ``what works and
what doesn't'' in order to carry out the continuous stream of modifications
that must be made. This framework can be used to help compare the different
ideas in meaningful ways;  without standard points of reference it is nearly
impossible to assess the effectiveness of one given approach relative to
another.

As might be expected of a rapidly growing field, there is no established
procedure for coordinating code tests with code development. In Section 2, we
provide an overview of some of the current methodology being practiced to
measure accuracy and stability. This provides the background for the discussion
in Section 3 of our strategy in designing a series of code tests and
comparisons which isolate in an effective way a set of performance levels
necessary for successful simulation of black holes. In Section 4, we discuss
the design specifications of the first round of tests, constructed so that code
performance can be based on common output obtained from common input run on
common grids.  In Section 5 we summarize our work and discuss future
perspectives. Our goal is to raise issues, collect ideas, point out pitfalls,
document experiences and in general promote and stimulate work toward a better
understanding of what works, what doesn't and -- ultimately -- why.

We use the following conventions here: the spacetime metric has
signature $(-,+,+,+)$; we use geometric units $G=c=1$;
the Ricci tensor is $R_{ab} = {R_{acb}}^{c}$; the
extrinsic curvature is defined as $K_{ab} = - \frac{1}{2}{\cal L}_{n} h_{ab}$;
where $h$ is the induced 3-metric and $n$ is the future directed timelike unit
normal; this means that positive extrinsic curvature signifies collapse, and
negative extrinsic curvature expansion.

\section{Current methodology}

\begin{table}[htbp]
  \begin{center}
    \begin{tabular}[c]{|l|l|l|c|}
 \hline
  Test & Comments & Formulation & Reference  \\ \hline
  Slicings and perturbations  & Harmonic gauge & ADM & \cite{Szilagyi02a}  \\ 
  of Minkowski space-time & Periodic gauge wave & KST / BM & \cite{Calabrese02a,Calabrese02b,Bona02a}  \\
  & Periodic gauge wave with shift & ADM / BM & \cite{Bona98b}  \\ 
  & Gaussian gauge pulse &  ADM / BM / BS &\cite{Balakrishna96a,Bona98b,Alcubierre99e}  \\ \hline
  Hyperboloidal gauges &
  Compactifications of Minkowski & CFE & \cite{Huebner99,Husa02b}  \\
  & Radiation of Brill type & CFE & \cite{Husa02b}  \\
  & Weak data & CFE &\cite{Huebner01,Husa02a}  \\
  & Asymptotically A3 & CFE, $\lambda$-CFE &\cite{Huebner99,Siebel01} \\ \hline
  Linearized solutions & Plane waves & ADM / BM &\cite{Bona98b,Anninos94d,Anninos96b} \\
  & Quadrupolar waves & ADM / BM &\cite{Bona98b,Allen97a,Anninos94d}  \\
  & Teukolsky waves & ADM / BM / BSSN &\cite{Anninos94d,Anninos96b,Abrahams97a,Baumgarte99} \\ 
  & Scattering off Schwarzschild & ADM & \cite{Rezzolla97a} \\ 
  & TT waves & SN & \cite{Shibata95} \\ \hline
  Nonlinear plane waves & geodesic/densitized geodesic slicing &
  Ashtekar / $\lambda$-Ashtekar &\cite{Shinkai00b,Yoneda01a} \\ \hline
  Gowdy Spacetimes & Expanding polarized & ADM & \cite{New98}  \\
  & Collapsing polarized & ADM & \cite{Garfinkle02} \\ \hline
  Robertson-Walker & $k=0$ (De Sitter) & BSSN &\cite{Vulcanov01}  \\ 
  &  $k=0$ (De Sitter) with cosmological constant & BSSN &\cite{Vulcanov01}
   \\ \hline 
  Kasner & & BSSN &\cite{Vulcanov01}  \\ \hline
  Brill Waves &
  Maximal and 1+log slicing & BSSN & \cite{Alcubierre99d}  \\ \hline
  Schwarzschild & Isotropic, geodesic slicing & ADM / BM &
  \cite{Bona98b,Bruegmann96,Anninos94c}  \\
  & Isotropic, algebraic slicing & ADM / BM &\cite{Bona98b,Arbona99,Anninos94c} \\
  & Isotropic, maximal slicing & ADM / BM &\cite{Bona98b,Anninos94c} \\
  &Kerr-Schild (ingoing Eddington-Finkelstein) & \begin{tabular}{l}
  adjADM/ KST / BSSN /  \\ LS / adjBSSN \end{tabular}&
  \cite{Kelly01,Kidder01a,Laguna02,Alcubierre00a,Yo02a} \\
  &Painlev\'e-Gullstrand & adjADM / KST & \cite{Kelly01,Kidder01a} \\
  &Harmonic time slicing & KST &\cite{Kidder01a} \\ \hline
  Distorted Schwarzschild &  & ADM / BM / BSSN & \cite{Camarda97b,Baker99a,Allen97a,Allen98a}
  \\ \hline 
  Kerr &
  Kerr-Schild & KST / adjBSSN & \cite{Kidder01a,Yo02a} \\ \hline
  Distorted Kerr &  algebraic slicing & BSSN & \cite{Alcubierre02a} \\ \hline
  Misner Black Holes &
  Maximal slicing & ADM / BM / BSSN &
  \cite{Anninos93b,Anninos94b,Price94b,Anninos96c,Alcubierre99d,Baker00b}
  \\ \hline

    \end{tabular}
    \caption{Solutions of the vacuum Einstein equations that have been used to
      validate Cauchy codes or formulations in the literature.
      The abbreviations of the formulations
      are following: 
          ADM=Arnowitt-Deser-Misner \cite{Arnowitt62,York79},
          adjADM=adjusted ADM \cite{Kelly01,Shinkai02},
          BM=Bona-Mass\'o \cite{Bona94b}, 
          BSSN=Baumgarte-Shapiro-Shibata-Nakamura \cite{Baumgarte99},
          adjBSSN=adjusted BSSN \cite{Yoneda02a},
          CFE=Conformal field eqs \cite{Friedrich81a,Friedrich81b},
          SN=Shibata-Nakamura \cite{Nakamura87,Nakamura89,Shibata95},
          KST=Kidder-Scheel-Teukolsky \cite{Kidder01a}, 
          LS=Laguna-Shoemaker \cite{Laguna02}, 
          Ashtekar \cite{Yoneda99a,Yoneda00a}, and 
          $\lambda$ = $\lambda$-system \cite{Brodbeck98,Shinkai99b}. 
    }
    \label{tab:exactsolns}
  \end{center}
\end{table}

As one of the primary aims of numerical relativity is to study spacetimes
containing black holes, an important test of
any code is its accuracy in simulating a black hole spacetime for which
the solution is known analytically, such as Schwarzschild. However it is
simpler to initially validate a code using spacetimes without
singularities. This is illustrated by the list in
Table~\ref{tab:exactsolns} of some of the solutions previously used in
the literature to validate 3+1 Cauchy vacuum codes. (A different set of
testbeds has been used for characteristic codes~\cite{Bishop97b,Husa02c}.) The
majority of tests have either used very weak field spacetimes, such as
Minkowski space in various slicings or linearized waves, or used
analytically known black hole spacetimes.

Most tests in Table~\ref{tab:exactsolns} were developed for a particular
application of a particular code. An attempt to define a general test suite for
numerical relativity was considered quite early in~\cite{Centrella86b}. There
the authors proposed 22 separate tests, 19 of which have an analytic solution
(sometimes only in the Newtonian or linearized regime) or have quantities that
can be calculated analytically. The emphasis in~\cite{Centrella86b} is strongly
toward relativistic hydrodynamics and only 5 of these tests can be applied to a
vacuum code. These 5 tests focus on linearized and weak gravitational waves.
The only strong field test is the simulation of a Brill wave, which does not
have an analytic solution but if the amplitude is insufficient to form a black
hole then the wave should disperse and the final radiated energy should equal
the initial mass-energy.

Two points should be noted about the tests originally suggested
in~\cite{Centrella86b}. First, none consider singularity formation or black
holes. Since then, the binary black hole problem has become the primary focus
of vacuum relativity codes and many subsequent tests have concentrated on
single black holes, such as Schwarzschild (in various slicings) or Kerr.
Perturbative solutions for distorted single black holes and their quasinormal
modes also provide important checks. Second, all the tests
in~\cite{Centrella86b} were intended to test code accuracy. None of them was
specifically intended to test stability. However stability has been the major
problem for three dimensional relativity codes. Although the same analytic
solutions could be used to study the stability of a code or of a formulation,
the design of the test and the analysis of the results must be changed.

The standard approach in the field has not been to use a wide range of tests as
suggested by~\cite{Centrella86b}. The tests in Table~\ref{tab:exactsolns} were
usually tailored to the specific problem of interest. For those papers
interested in wave extraction, popular tests have included slicings of
Minkowski, linearized and weak gravitational waves. Papers interested in black
hole physics have used static or stationary black holes in various slicings.
Evolutions of distorted black
holes~\cite{Camarda97b,Baker99a,Allen97a,Allen98a,Alcubierre02a} or colliding
black holes~\cite{Price94b,Anninos96c,Alcubierre99d,Baker00b} have also been
used to provide benchmarks for fully nonlinear codes. Papers investigating
instabilities of various formulations or boundary conditions have tended to use
slicings of Minkowski space~\cite{Calabrese02a}, weakly constraint violating
data~\cite{Szilagyi00a,Szilagyi02a} or linearized waves, although black hole
solutions have also been used~\cite{Yo02a}. 

Many of these papers have concentrated on the \textit{convergence} of the
numerical code and have assumed that convergence is sufficient for a physically
valid solution, even without a proof of the well-posedness of the continuum
equations. The problems with this approach have been emphasized by various
authors~\cite{Calabrese02a,Szilagyi00a,Szilagyi02a}. Many formulations in use
are either not well-posed or are not known to be well-posed with the gauge
conditions or boundary conditions in practice. For example, many of the
boundary conditions in use are inconsistent with the constraints and the
evolution will converge to a valid solution only for the short time  when it is
causally disconnected from the boundary. As shown by Kreiss and
Oliger~\cite{Kreiss73} and illustrated in the context of numerical
relativity~\cite{Calabrese02a,Szilagyi00a}, such gauge or boundary
inconsistencies can lead to exponentially growing modes that may not appear in
tests run at low resolutions or short time scales.

Choptuik~\cite{Choptuik91,Choptuik02} has emphasized that if the
continuum equations governing the evolution system and the constraints
are well-posed and if the differencing scheme is consistent then some
form of a Lax equivalence theorem should be expected to apply, i.e.\ 
that convergence and stability should be equivalent. Even in the
absence of an exact solution, convergence can be checked in the Cauchy
convergence sense by Richardson extrapolation and consistency can be
checked by ``independent residual evaluation'' using an alternative
discretization of the equations obtained by symbolic algebra
techniques (see e.g.~\cite{Miller00}). These methods give great
confidence in the correctness of the numerical solution on the
assumption that the underlying system of equations is well-posed. But
that assumption lies beyond the analytic understanding of present day
codes used to tackle the binary black hole problem.

\bigskip
 
\section{Design of standardized tests}

Tests for numerical relativity should ideally satisfy a number of important
properties. They should be broadly applicable, in the sense that they can be
executed within a broad range of formulations. They should produce data which
are unambiguous. They should specify free quantities such as the gauge. They
should output variables which are independent of the details of the model and
can be used for comparison. Finally, the tests and their output should provide
some insight into both the characteristics of the code on which they are run and
the comparative performance of other codes run on the same
problem.

The first of these issues, universality, is not difficult to satisfy within a
given class of codes. For instance, we can consider the class of ``$3+1$''
codes which evolve spacelike surfaces on a finite domain with periodic boundary
conditions. For most formulations it should be a straightforward technical
problem  to express some initial data set in variables appropriate for a given
formulation, and to construct the desired output variables at each time step.
Even within this class of codes, however, details of implementation might
restrict the set of tests which a particular code could run. For instance,
codes which are fixed to run on a spherical domain might be difficult to test
directly using data specified on a cube with periodic boundaries. 

More serious difficulties arise in cross-comparing between classes of codes
based upon entirely different formulations. Direct comparison of a finite
domain code with a compactified characteristic code or with a Cauchy code based
upon hyperboloidal time slices extending to null infinity would be impossible,
given the difficulties in defining equivalent initial data. However, for a
Cauchy code run on a large spatial domain, one could imagine comparing
physically motivated quantities, such as the extracted radiation, with the
results of a compactified code.

Of course, each test must specify an appropriate set of output quantities for
analysis and comparison. It is common practice to judge the performance of a
relativity code by measuring the degree of violation of the constraint
equations.  A properly working code should satisfy each of its constraints at
each time-step. Constraint violation is a clear indication of a problem and
usually the first indicator of growing modes which will eventually kill a
simulation. On the other hand, a code which accurately preserves the
constraints, perhaps by enforcing them during the evolution, might suffer other
losses of accuracy which produce a numerically generated spacetime that is
unphysical, even though it lies on the constraint hypersurface (for an example,
see~\cite{Siebel01}). Solutions to the constraint equations are not necessarily
connected via the evolution equations (e.g., a Schwarzschild spacetime should
not evolve into Minkowski space, even though both will satisfy constraints
perfectly!). Indeed, mixing constraint equations with the evolution equations
can have great effect on the numerical results~\cite{Yo02a}. For these reasons,
it is important not to examine the constraints in isolation. Similarly, the
length of time a code runs before it ``crashes'' is not an appropriate
criterion of quality unless accompanied by some indication of how accurately
the code reproduces the intended physics, for example the constant amplitude
and phase of a wave in the linear regime propagating inside a box with periodic
boundaries.  

Ideally, one would like to compare the convergence of variables against exact
solutions where the answer is known. These are the most unambiguous tests and
the most important for debugging a code. The exact solution provides an
explicit choice of lapse and shift which can be adopted by any code through the
use of gauge source functions. Furthermore, the exact solution provides
explicit boundary data. 

However, exact solutions for complex physical problems in relativity are
scarce. We need additional criteria by which to judge whether numerically
generated spacetimes are ``correct''. For this one could include physical
criteria, such as the energy balance between mass and radiation loss (for
example, see~\cite{Brandt94c,Alcubierre00b}). One could check some expected
physical behavior of an evolution, such as the dispersion of a small amplitude
Brill wave to evolve toward Minkowski spacetime, or the gravitational collapse
of a high amplitude wave~\cite{Alcubierre99b}. Unfortunately, such comparisons
between codes can be misleading because of coordinate singularities produced by
choice of gauge or by improper choice of boundary condition. For instance, the
two spacetimes generated with the same initial data by two codes based upon
harmonic time slicing might exhibit qualitatively different regularity
properties because of different choices of shift. In such a case, the resulting
difference in time slicings even complicates comparisons based upon curvature
invariants because of the difficulty in identifying the same spacetime point in
two different simulations. The same is true of spacetimes generated with the
same initial data and gauge conditions but with different boundary conditions. 
Brill wave initial data which disperses to Minkowski space with a Sommerfeld
boundary condition might collapse to a singularity with a Dirichlet boundary
condition. 

In the absence of exact solutions, convergence can be measured in the sense of
Cauchy convergence. One might consider that a set of tests could be based upon
the agreement between a number of independent convergent codes. However,
caution needs to be taken with this approach since a common error could lead to
acceptance of an incorrect numerical solution which then might systematically
bias the modification of later codes to reproduce the accepted solution.
Although convergence criteria are an important measure of the quality of any
simulation, systematic problems can lead a code to converge to a physically
irrelevant solution. Furthermore, agreement between codes is only
straightforward when they use the same choice of lapse and shift, which is not
always possible. Convergence tests should be accompanied by some expectation of
how physical variables should behave. Initial data which have some physical
interpretation, such as propagating waves, provide useful tests since they
provide some picture of how the fields should behave. Any numerical relativity
code should be able to reproduce such expected physical behavior if it is to be
trusted when unexpected phenomena are encountered.

Ideally, tests should  be ``simple'' as well as physical. For a simple enough
test, some properties of a correct solution can be developed even if an exact
solution is not known. Particular advantages and disadvantages of a code can be
more readily isolated using simple tests for which the behavior of individual
variables is understood. Of special importance, a simple test can be more
readily implemented by a broad set of codes.

A characteristic feature of asymptotically flat systems is that energy gets
lost through radiation. Correspondingly, the numerical treatment of such
systems typically leads to an initial boundary value problem. However, in order
to create a test suite which allows separation of problems associated with the
boundary treatment from other problems, we believe it is fruitful to also
consider tests with periodic boundaries. The use of periodic boundary
conditions in testing numerical codes is a standard technique in computational
science. In the context of general relativity, however, such tests have to be
set up and interpreted with great care, since periodic boundaries create the
cosmological context of a compact spatial manifold with the topology of a
3-torus. The resulting physics is qualitatively very different from
asymptotically flat solutions. An example is the nonlinear stability of
Minkowski spacetime, on which -- at least implicitly -- many numerical
stability tests rely: For weak asymptotically flat data, it has been proved
that perturbations decay to Minkowski
spacetime~\cite{Christodoulou93,Friedrich86}. This result does {\em not} hold
in the context of periodic boundaries! For two Killing vectors, it has been
shown~\cite{Andreasson02} that all initial data which are not flat fall into
two classes which are related to each other by time reversal. Making the
standard cosmological choice of time direction, all non-flat data have a
crushing singularity in the past and exist globally in a certain sense in the
future. In particular, the spacetime can be covered by constant mean curvature
hypersurfaces whose mean curvature goes to zero in the
future~\cite{Andreasson02, Jurke03}. In the simulation of such a spacetime,
although the initial data analytically implies expansion, discretization error
or constraint violation can drive the numerically perturbed spacetime toward a
singularity in the future.

When $trK > 0$ (in the present conventions) everywhere on a time slice, the
pitfalls are easy to recognize. In that case, the singularity theorems (see
e.g.~Refs.~\cite{Hawking73a} or \cite{Wald84}) imply that a singularity must
form in finite proper time -- even if the initial data are arbitrarily close to
locally Minkowski data. Apart from such physical singularities, focusing
effects can also create gauge singularities which could also lead to a code
crash. Even in the $trK < 0$ case where the spacetime undergoes infinite
expansion, the code can crash due to choice of evolution variables that become
infinite.

These observations introduce various complications for the setup and
interpretation of numerical tests with periodic boundaries. For ``linearized
waves'' with initial data close to Minkowski data, constraint violation or
other source of error might lead to either expansion or collapse, with
different codes exhibiting different behavior. For Minkowski data in
nonstandard coordinates the situation is similar: Individual runs should be
expected to show instabilities but with grid refinement the runs should show
convergence to Minkowski spacetime. For situations with $trK > 0$ on a whole
time slice all codes should exhibit eventual collapse but in other cases the
qualitative behavior may be code dependent. In order to interpret the results
of a simulation it is thus vital to probe the underlying dynamics in terms of
expansion versus collapse -- in particular since many gauge conditions for the
lapse involve $trK$. For this purpose, it is important to monitor such
variables as $trK$, the eigenvalues of $K_{ab}$, $\det(g)$ and the total volume
or proper time.

Since any given test will satisfy only a few of the desired criteria, some
balance between redundancy and economy is needed. We can envisage a hierarchy
of tests, starting from evolving flat space under various gauge assumptions, to
linearized and then non-linear waves, to perturbations of a stationary black
hole and then eventually to highly non-linear, dynamic black holes. Each
successive test should introduce a new feature for which code performance can
be isolated. The major goal of numerical relativity is the simulation of binary
black holes. This requires special techniques, such as singularity excision,
which by themselves are an extreme test for any code and can obscure the
precise source of an instability arising in such a strong field regime. A
standardized test suite should lead up to the binary problem through models of
static, moving and perturbed black holes.

It is not our intention in this paper to present the specifics of a complete
test suite. We concentrate here on an initial round of simple tests which serve
to highlight certain important characteristics of the codes represented at the
Mexico workshop and which can be readily performed with most codes. Even in
trial implementations of these {\em simple} tests we have found {\em
complications} that warn us that continuous feedback between design and
experiment is absolutely necessary in developing a full test suite. The four
tests in this initial round are: (i) the robust stability test, (ii) the gauge
wave test, (iii) the linearized wave test and (iv) the Gowdy wave test.

Robust stability~\cite{Szilagyi00a} is a particular good example of a testbed
satisfying the above criteria. Random constraint violating initial data in the
linearized regime is used to simulate unavoidable machine error. It can be
universally applied, since any code can run perturbations of Minkowski space,
and it is very efficient at revealing unstable modes, since at early times they
are not hidden beneath a larger signal. Even in a convergence test based upon a
non-singular solution, smoother truncation error dominates, at least until very
late times, unless the resolution is very high. In stage I of the testbed,
which we include in our first round of tests, evolution is carried out on a
3-torus (equivalent to periodic boundary conditions) in order to isolate
problems which are independent of the boundary condition. In Stage II, one
dimension of the 3-torus is opened up to form a 2-torus with plane boundaries,
where random boundary data is applied at all times. This tests for stability in
the presence of a {\em smooth boundary}. Boundary conditions which depend upon
the direction of the outer normal, such as Neumann or Sommerfeld conditions,
are best tested first with smooth boundaries in order to isolate problems. In
Stage III,  random data is applied at all faces of a cubic boundary, the 
common choice of outer boundary in simulating an {\em isolated system}. This
tests for stability in the presence of {\em edges and corners} on the boundary.
The test can be extended to Stage IV in which a spherical boundary is cut out
of a Cartesian grid. These tests provide an efficient way to cull out unstable
algorithms as a precursor to more time consuming convergence tests. The
performance can be monitored by outputting the value of the Hamiltonian
constraint.

The gauge wave testbed also very aptly fits our criteria. An exact wave-like
solution is constructed by carrying out a coordinate transformation on
Minkowski space. The solution can be carried out on a 3-torus, by matching the
wavelength to the size of the evolution domain, or it can be carried out in the
presence of boundaries. Since the gauge choice and boundary data are explicitly
known, it is easy to carry out code comparisons. The evolution can be performed
in the weak field regime or in the extreme strong field regime which borders on
creating a coordinate singularity. Knowledge of a non-singular exact solution
allows any instability to be attributed to code performance. Long, high
resolution evolutions can be performed. Convergence criteria for the numerical
solution can be easily incorporated into the test. In our first round of tests,
we simulate a gauge wave of moderately non-linear amplitude   propagating
either parallel or diagonal to the generators of a 3-torus. 

The linearized wave testbed uses a solution to the linearized
Einstein equations and is complimentary to the first two. Run in the
linearized regime, it provides an effectively exact solution of physical
importance which can be used to check the amplitude and phase of a 
gravitational wave as it propagates on the 3-torus. If this test were run with
a higher amplitude, so that nonlinear effects are not lost in machine noise, the
constraint violation in the initial linearized data would introduce
a complication. The tendency toward gravitational collapse could also amplify
numerical error or the effect of a bad choice of gauge, e.g.\ the focusing
effect inherent in Gaussian coordinates. A maximal slicing gauge cannot be used
to avoid such problems, because, for any nontrivial solution, it would be
inconsistent with periodic boundary conditions. While in the asymptotically
flat context it is typical to use maximal initial hypersurfaces, it is known
that a maximal Cauchy surface of $T^3$ topology has to correspond to locally
Minkowski data (see e.g.~Ref.~\cite{Schoen79} and pages 2-3 of
Ref.~\cite{Anderson00}). This result is connected to the fact that $T^3$ does
not admit metrics of positive  Yamabe number, with the Yamabe number of flat
$T^3$ being zero.  In order to avoid these complications, our nonlinear wave
tests, which complete the present round of testbeds,
are based upon polarized Gowdy spacetimes. 
These spacetimes provide a family of exact solutions describing an
expanding universe containing plane polarized gravitational
waves~\cite{Gowdy71} and a clear physical
picture to test against the results of a simulation.
We will carry out this test both in the expanding
and collapsing time directions, which yields physically very different
situations with potentially different mechanisms to trigger instabilities.

Except for the robust stability test, the remaining testbeds are based upon
solutions with translational symmetry along one or two coordinate axes. By
using the minimal required number of grid points along the symmetry axes, this
allows the tests to be run with several grid sizes without exorbitant
computational expense. Such tests could also be run with either 1D or 2D codes.
In order to check that a 3D code is not taking undue advantage of the symmetry,
the initial data can be superimposed with random noise as in the robust
stability test. For purposes of economy we do not officially include this as
part of the testbeds, but it is a useful practice which can completely alter
the performance of a code.

\section{test case specification}

We will specify the physical properties of each testbed by providing
the complete 4-metric of the spacetime, or if this is not possible,
the initial Cauchy data and choice of gauge for the evolution. In all
cases, we will give the Cauchy data, i.e.~the 3-metric and extrinsic
curvature, in a Cartesian coordinate system appropriate for
3-dimensional evolution. The physical domain is a cube
which, in this first round of tests with periodic boundary conditions,
represents a 3-torus.

In order for uniform comparison the series of four tests should be run
using a second order iterative-Crank-Nicholson algorithm with two
iterations (in the notation of~\cite{Teukolsky00}) with second order
accurate finite differencing in space. There may be codes that cannot
implement this type of numerical method. Similarly, a particular code
may run better with an alternative numerical method such as a
Runge-Kutta time integrator or a pseudo-spectral method. In such cases
the relative performance of the code for these tests still offers a
useful comparison, provided all parameters (such as the amount of
artificial dissipation) are held constant over the four tests.
However, for a quantitative cross comparison of codes it is best to
provide results from a standard numerical method. Second order
iterative-Crank-Nicholson is chosen for simplicity.

\begin{figure}
  \includegraphics[width=30pc]{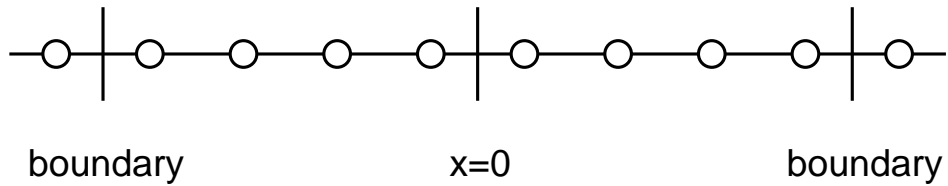}
  \caption{Grid points (in this case n=8) along a given axis are chosen to
    straddle both $x=0$ and
    the identified boundaries. An arbitrary number of ghost-zone
    points beyond the boundaries can be used in implementing
    periodic boundary conditions.}
\label{fig:periodic}
\end{figure}

The simulation domain for each test will generally be a cube of side $d$,
equal to one wavelength with periodic boundary conditions. The grids are
set up to extend an equal distance in the positive and negative
directions along each axis. As depicted in Fig.~\ref{fig:periodic}, the
``boundaries'', which are identified in the 3-torus picture, are located
a half step from the first and last grid points along each axis. The
resolution in a direction $i$ is given by $\Delta x^i = d/n^i$. The
number of grid points $n^i$ should be sufficient to resolve features of
the initial data in the given direction. Even though we are running
three-dimensional codes, for tests with only one-dimensional features it
is considerably more efficient to restrict the grid such that $n^i$ is
small in the trivial directions.  As an example, for a wave propagating
in the $x$-direction we use the minimum number of grid points in the 
trivial $y$ and $z$ directions that allow for non-trivial numerical
second derivatives. For standard second order finite differencing this
implies that we use 3 points in those directions. 

The size of the timestep $dt$ is given in terms of the grid
size $dx$ and chosen to lie within the CFL limit for an explicit evolution
algorithm. We foresee the possibility of codes for which
this would be inappropriate and for which some equivalent choice of
time step would have to be made. A final time $T$, and intermediate
times for data output, are specified for each test. The time $T$ is
chosen to incorporate all useful features of the test without
prohibitive computational expense. The output times should be
appropriately modified for codes that crash before time $T$.  

The output quantities are chosen with either some physical or numerical
motivation in mind. Both quantitative and qualitative comparisons are
used. Some examples, based upon representative versions of evolution
codes, are provided below for purposes of illustration. We specify a
minimum list of output quantities. Other quantities that we do not
include here but that may be of interest for a specific code include the
Fourier transform of differences between the numerical and exact
solutions as in~\cite{Calabrese02a}, curvature invariants to detect
deviations from flat space, and proper time integrated along observer
world lines. Each group should of course also output any additional
variables which are essential to the performance of their particular
formulation.


\subsection{Robust stability testbed}

The robust stability testbed~\cite{Szilagyi00a} efficiently reveals
exponentially growing modes which otherwise might be masked beneath a
strong initial signal for a considerable evolution time. It is based
upon small random perturbations of Minkowski space. The initial data
consists of random numbers $\epsilon$ applied as a perturbation at
each grid point to every code variable requiring initialization. For
example, the initial 3-metric is initialized as
$h_{ij}=\delta_{ij}+\epsilon_{ij}$, where the $\epsilon_{ij}$ are
independent random numbers. The range of the random numbers ensures
that $\epsilon^2$ effects are below roundoff accuracy so that the
evolution remains in the linear domain unless instabilities arise.

\begin{figure}
  \includegraphics[width=40pc]{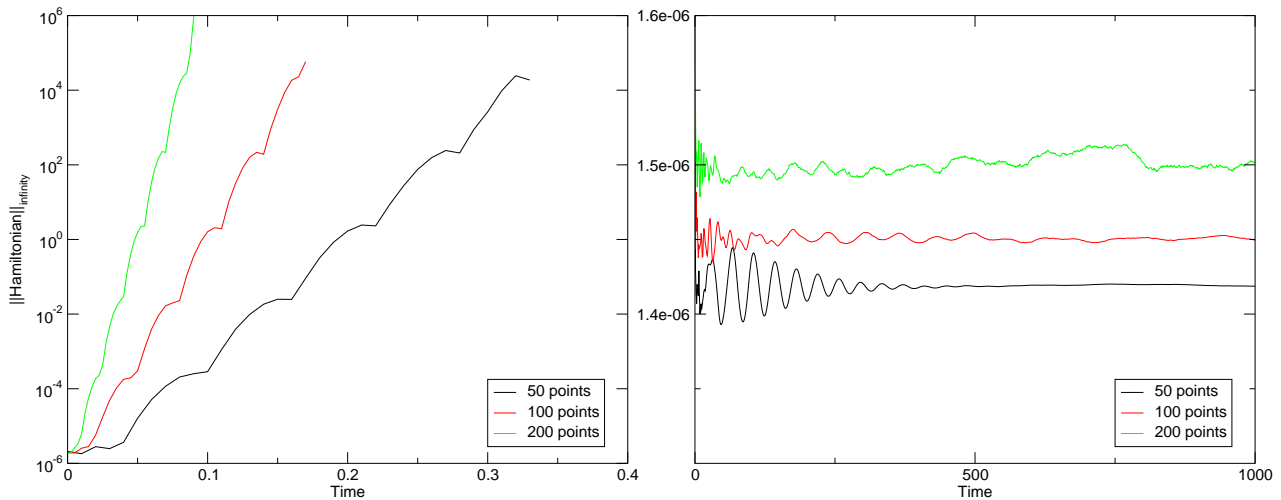}
  \caption{Left: the robust stability test applied to the standard
    formulation of the ADM equations. It is clear that there is an
    exponentially growing mode and that the growth rate of the mode
    depends on resolution. Right: the robust stability test applied to
    the BSSN formulation. The violation of the constraint is
    approximately constant even after 1000 crossing times. In both
    cases the harmonic gauge was used. Note the differences in the
    axes.}
  \label{fig:RobustStability_ADMvsBSSN}
\end{figure}

For economy,  we fix the following parameters:
\begin{itemize}
\item Simulation domain: $x \in [-0.5, +0.5]$
\item Grid: $x_n = -0.5 + (n-\frac{1}{2}) dx, \quad n=1\ldots 50\rho,
  \quad dx=dy=dz=1/(50\rho), \quad \rho \in \mathbb{Z}$
\item Time step: $dt = dx/2 = 0.01 / \rho$
\end{itemize}
The parameter $\rho$ allows for refinement testing, which clarifies
the results as can be seen in
Figure~\ref{fig:RobustStability_ADMvsBSSN}. The values used in that
test were $\rho=1,2,4$. We use the minimum number of grid points in
the $y,z$ directions that allow for non-trivial numerical second
derivatives -- this means we carry out this test in a long channel rather
than a cube. For standard second order finite differencing this
implies that we use 3 points in the $y$ and $z$ directions.

The amplitude of the random noise should be scaled with the grid
spacing as
\begin{equation}
  \label{eq:RobustStabilityAmplitude}
  \epsilon \in (-10^{-10}/\rho^2,+10^{-10}/\rho^2).
\end{equation}
This ensures that the norm of the Hamiltonian constraint violation in
the initial data will be (on average) the same for different values of
the refinement factor $\rho$. This means that in the continuum limit
we will have a solution that is not a solution of the Einstein
equations but ``close'' to one. This would be the case in a real
numerical evolution where machine precision takes the place of
$\epsilon$. If a code cannot stably evolve the random noise then it
will be unable to evolve a real initial data set.

The test should be run for a time of $T=1000$ (corresponding to $1000$
crossing times) or until the code crashes. Performance is monitored by
outputting the $L_\infty$ norm of the Hamiltonian constraint once per
crossing time, i.e.\ at $t=0,1,2,3,...$.  Because the initial data
violates the constraints, any instability can be expected to lead to
an exponential growth of the Hamiltonian unless enforcement of the
Hamiltonian constraint were built into the evolution algorithm. As an
example, Figure~\ref{fig:RobustStability_ADMvsBSSN} shows the
performance of standard ADM and BSSN codes and illustrates the
efficacy of this test in revealing unstable codes.


\subsection{Gauge wave testbed}

These tests look at the ability of formulations to
handle gauge dynamics. This is done by considering flat Minkowski
space in a slicing where the 3-metric $h_{ij}$ is time dependent.
Such gauge waves have been considered before, notably
by~\cite{Alcubierre97a} and \cite{Calabrese02a}.

We have considered different profiles for gauge waves. For the purpose of this
paper we focus on the case of a propagating gauge sine wave.
This specific test was used in comparing systems with different hyperbolicity
properties in~\cite{Calabrese02a}. 

The 4-metric is obtained from the Minkowski metric
$ds^2 =- d\hat t^2+d\hat x^2+d\hat y^2+d\hat z^2$ by the coordinate
transformation 
\begin{equation}
  \label{eq:GaugeWave1}
  \begin{array}[c]{r c l}
   \hat t&=& t + \frac {Ad}{4\pi}\cos \left( \frac{2 \pi (x - t)}{d} \right), 
      \\
   \hat x&=& x - \frac {Ad}{4\pi}\cos \left( \frac{2 \pi (x - t)}{d} \right), 
      \\
     \hat y&=& y,    \\ 
     \hat z&=& z
  \end{array}
\end{equation}
where $d$ is the size of the evolution domain. This leads to the 
4-metric
\begin{equation}
  \label{eq:flatgaugewave4metric}
  ds^2=-H dt^2 +Hdx^2+dy^2+dz^2,
\end{equation}
where
\begin{equation}
  \label{eq:flatgaugewaveHfn}
  H = H(x-t)=1 + A \sin \left( \frac{2 \pi (x - t)}{d} \right),
\end{equation}
which describes a sinusoidal gauge wave of amplitude $A$ propagating
along the $x$-axis. The extrinsic curvature is given by
\begin{eqnarray}
  \label{eq:flatgaugewaveK}
  K_{xx} &=& -\frac{\pi A}{d} \frac{ \cos \left( \frac{2 \pi (x - t)}{d}
  \right) }{ \sqrt{1 + A \sin \left( \frac{2 \pi (x - t)}{d} \right) }
  },\\
K_{ij} &=& 0 \qquad \textrm{ otherwise}.
\end{eqnarray}

Since this wave propagates along the $x$-axis and all derivatives are
zero in the $y$ and $z$ directions, the problem is essentially one
dimensional and can simplify the system dramatically
for certain formulations, as there is no finite-difference error in
the orthogonal directions. A simple coordinate transformation causes
the wave to propagate along a diagonal:
\begin{equation}
  \label{eq:flatgaugewave1to2d}
x = \frac{1}{\sqrt{2}}(x^\prime - y^\prime), \qquad
y = \frac{1}{\sqrt{2}}(x^\prime + y^\prime)
\end{equation}
The resulting metric is a function of 
\begin{equation}
\sin \left( \frac{2 \pi (x' - y' - t' \sqrt{2})}{d'} \right), 
\quad \textrm{where} \quad d' = d \sqrt{2} \, .
\end{equation}
Setting $d'$ to the size of the evolution domain in
the $x'$ and $y'$ directions gives periodicity along those directions.
This test should be run in both axis-aligned and diagonal
form.

As any evolution is a pure gauge effect, care must be taken in the
choice of lapse and shift to allow a direct comparison between
formulations. For example, the Bona-Masso family of gauges \cite{Bona94b}
\begin{equation}
  \label{eq:1+log}
  \partial_t \alpha = - \alpha^2 f(\alpha) K
\end{equation}
will propagate a gauge wave in direction $x^i$ with speed 
$\alpha \sqrt{f g^{ii}}$, which can be varied.
In contrast, maximal slicing would freeze any gauge pulse,
stopping it from propagating.

Note that the time coordinate $t$ in the metric
(\ref{eq:flatgaugewave4metric}) is harmonic, which corresponds to
\begin{equation}
  \label{eq:harmonic}
  f(\alpha) = 1.
\end{equation}
This gauge condition can easily be integrated to 
$\alpha = h(x^i) \sqrt(\det g)$. Also, in this case the gauge speed
is simply the speed of light.
To ensure that we can directly compare as many formulations as possible
we choose harmonic slicing to carry out the evolution for the gauge
wave test.
Different formulations and codes may demand different
implementations of harmonic time slicing for optimal performance.
The test should thus be implemented in a way that is analytically
compatible with  the metric (\ref{eq:flatgaugewave4metric}), which still
leaves significant freedom.

We run the gauge wave with amplitudes $A=10^{-1}$ and $A=10^{-2}$. We have
found that smaller amplitudes are quite simple for all codes whilst larger
amplitudes can cause numerical error to trigger gauge pathologies, such as the
formation of coordinate singularities, very quickly.

The specified wave has wavelength $d=1$ in the 1D simulation and
wavelength $d'=\sqrt{2}$ in the diagonal simulation. We find that 50 grid
points are sufficient to resolve the profile and therefore make the
following choices for the computational grid:
\begin{itemize}
\item Simulation domain:
\begin{center}
\begin{tabular}{rllll}
  1D:& $\quad  x \in [-0.5; +0.5],$ & $\quad y = 0,$ &$ \quad z = 0,$
  & $ \quad d=1$ \\
  diagonal: & $\quad x \in [-0.5; +0.5], $ & 
    $\quad y \in [-0.5; +0.5], $ & $\quad z=0,$ & $\quad d'=\sqrt{2}$
\end{tabular}
\end{center}
\item Grid: $x^i = -0.5 + (n-\frac{1}{2}) dx, \quad n=1\ldots 50\rho,
  \quad dx=1/(50\rho), \quad \rho \in \mathbb{Z}$
\item Time step: $dt = dx/4 = 0.005 / \rho$
\end{itemize}
The 1D evolution is carried out for $T=1000$ crossing times,
i.e.~$2\times10^5\rho$ time steps (or until the code crashes), with
output every 10 crossing times.  The 2D diagonal runs are carried out
for $T=100$, with output every crossing time. We run using
$\rho=1,2,4$.

\begin{figure}
  \includegraphics[width=14pc]{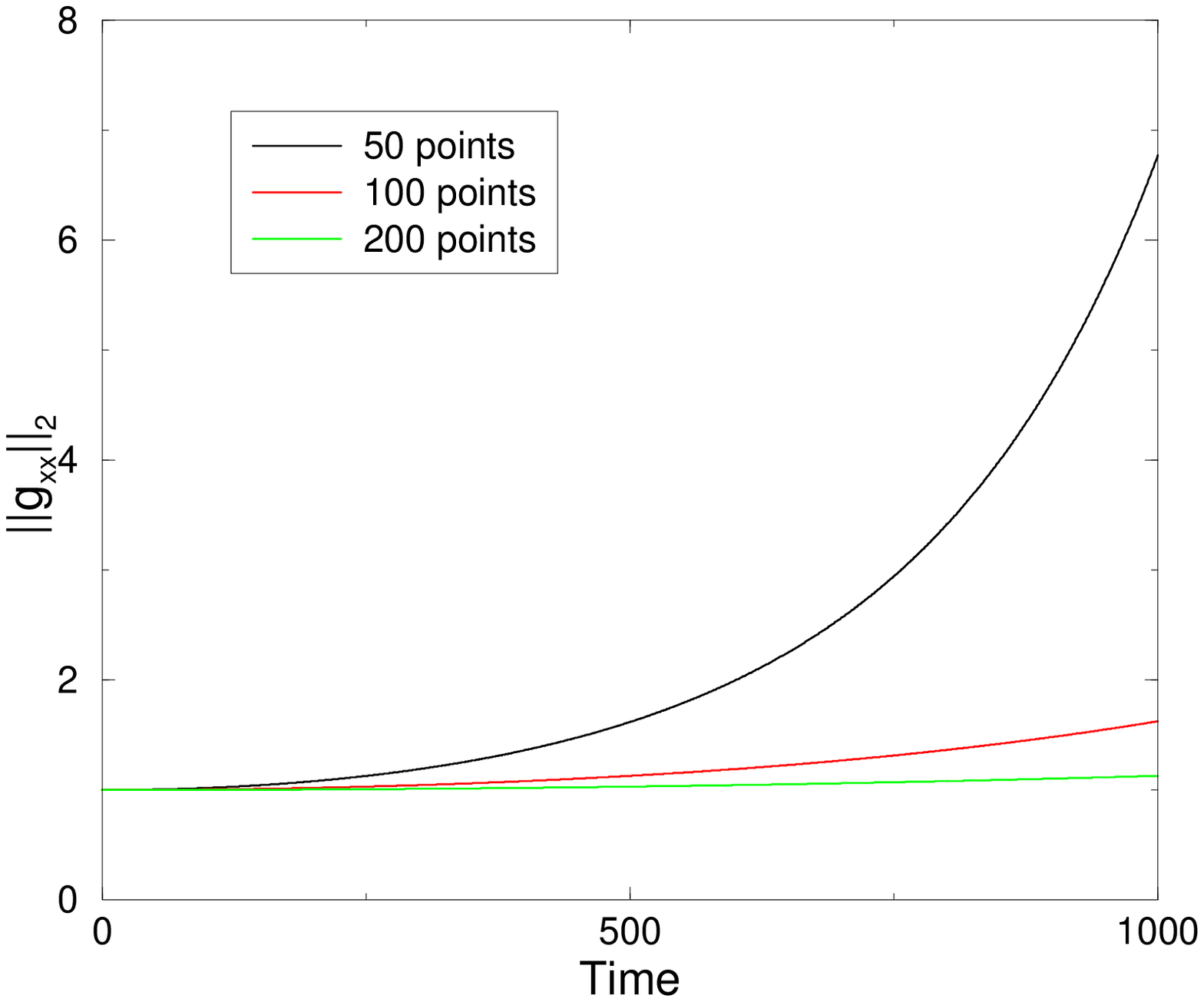}
  \includegraphics[width=14pc]{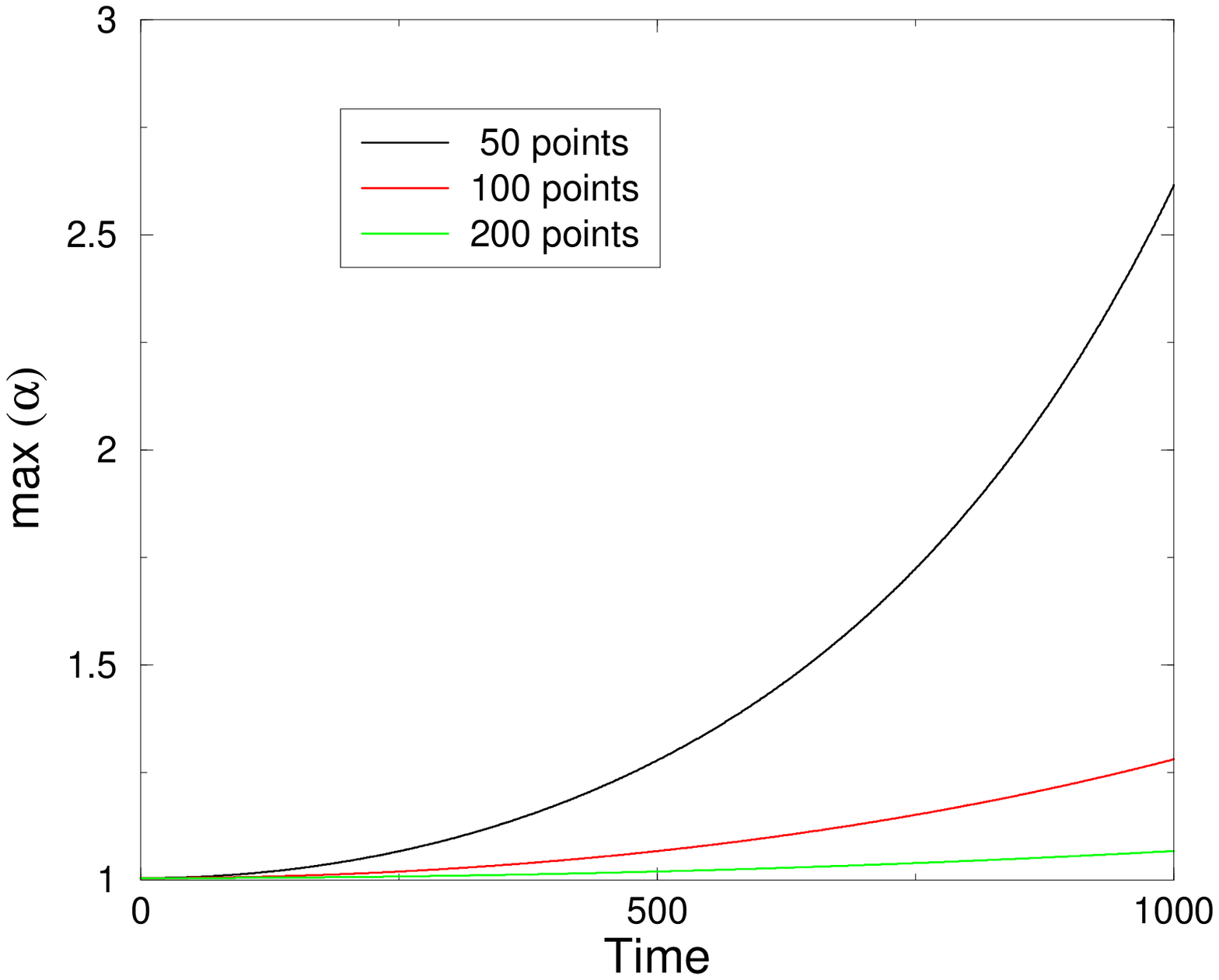}
  \includegraphics[width=14pc]{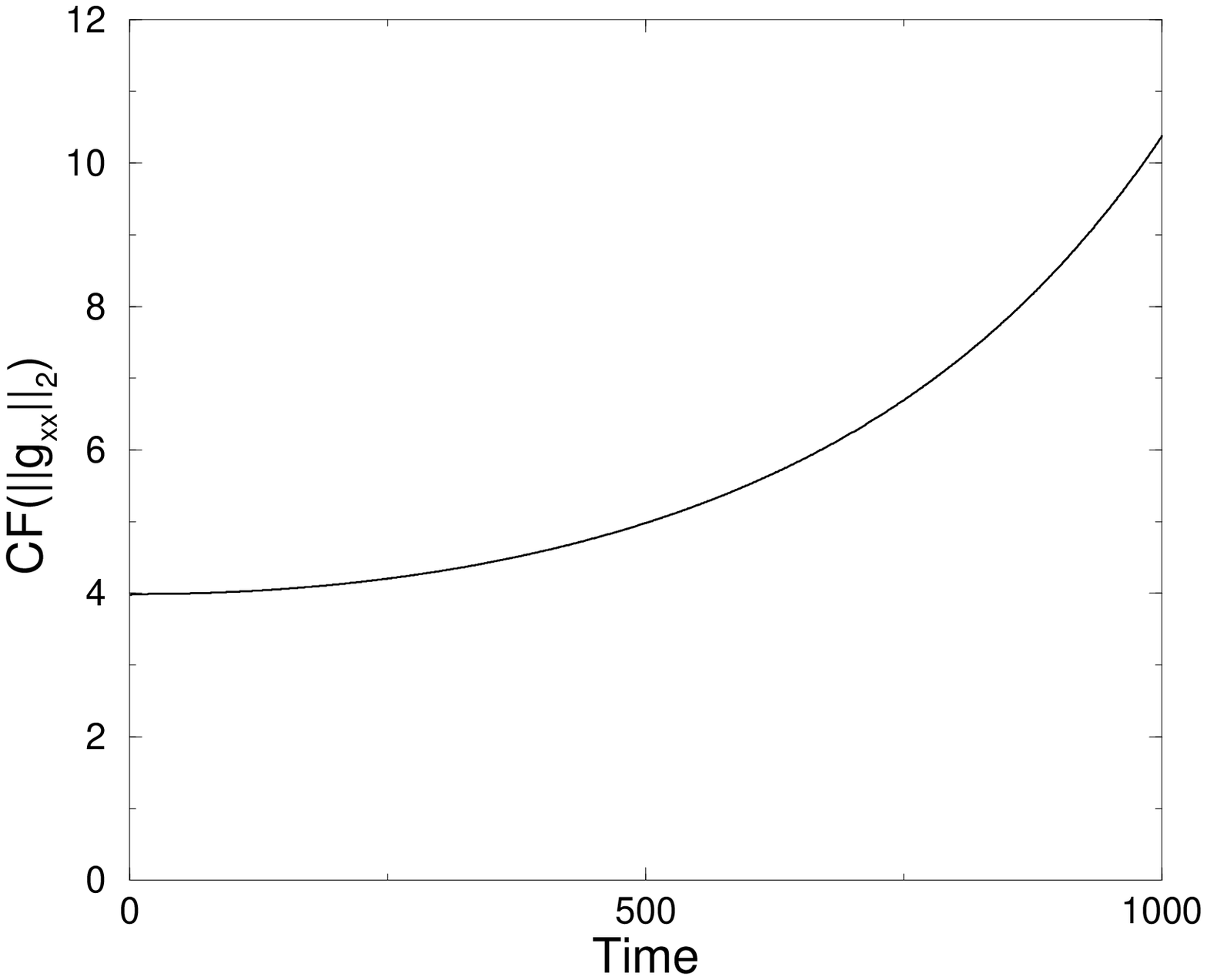}
  \caption{Results for the 1D gauge wave using the standard ADM formulation.
    The left hand plot shows $||g_{xx}||_2$; the central plot shows
    the maximum of the lapse $\alpha$, and the right hand plot shows
    the convergence factor of $g_{xx}$ calculated using the three
    resolutions. A value of $4$ would mark exact second order
    convergence.}
  \label{fig:GW_ADM_1D}
\end{figure}

We output the $L_{2}$, the maxima and minima, and
profiles along the $x$-axis through the center of the grid of
$g_{xx}$, $\alpha$, $tr(K)$, the Hamiltonian constraint and any other
independent constraints that arise in a nontrivial way in a particular
formulation. We also calculate the $L_{2}$-norm of the difference from
the exact solution for $g_{xx}$ and calculate the convergence factor.
Figure \ref{fig:GW_ADM_1D} provides examples of test output obtained
with a standard ADM code. These results do not show a problem with ADM,
but illustrate the expansion of the numerical space-time.  This is due
to the reasons mentioned in section III, stating that any non trivial
space-time with $T^3$ topology must have a singularity either in the
past or in the future. The behavior of the $L_2$-norm of $g_{xx}$
indicates how the volume element of the space behaves. It is also seen
that the evolution is convergent for a long time, nevertheless the higher 
order terms cause the deviation from convergence.

\subsection{Linear wave testbed}

This test checks the ability of a code to propagate the amplitude and phase
of a traveling gravitational wave. The test is run in the linear regime
where there are no complications due to the toroidal topology implicit in
periodic boundary conditions. It reveals effects of numerical dissipation
and other sources  of inaccuracy in the evolution algorithm. We note that
the evolution is meaningless once the accumulation of numerical error takes
it out of the linear regime.

The initial 3-metric and extrinsic curvature $K_{ij}$ are given by a
diagonal perturbation with components
\begin{equation}
  \label{eq:linearwave4metric}
  ds^2= - dt^2 + dx^2 + (1+b) \, dy^2 + (1-b)  \, dz^2,
\end{equation}
where 
\begin{equation}
   b  =  A \sin \left( \frac{2 \pi  (x-t)}{d}\right)
\end{equation}
for a linearized plane wave traveling in the $x$-direction.
Here $d$ is the linear size of the propagation domain, and the
metric is written here in Gauss coordinates, i.e.\ with lapse
$\alpha=1$ and shift $\beta^i=0$.
The nontrivial components of extrinsic curvature are then
\begin{equation}
  \label{eq:linearwave4extcurv}
   K_{yy} =   \frac{1}{2} \partial_t b, \quad
   K_{zz} = - \frac{1}{2} \partial_t b.
\end{equation}

As in the case of the gauge wave, by the simple coordinate
transformation 
\begin{equation}
x = \frac{1}{\sqrt{2}}(y^\prime + x^\prime), \qquad
y = \frac{1}{\sqrt{2}}(y^\prime - x^\prime)
\end{equation}
the propagation direction can be aligned with a diagonal.
Setting $d' = d \sqrt{2}$ to the size of the evolution 
domain in the $x'$, $y'$ directions gives periodicity 
along those directions.

The amplitude of the wave is chosen as $A = 10^{-8}$, such that
quadratic terms are of the order of numerical roundoff. Larger
amplitudes mean that the solution does not stay in the linear regime
sufficiently long.

The geometry of the grid is chosen identical to the 1D gauge wave
test, with $d=1$ in the 1D case and $d'=\sqrt{2}$ in the diagonal case:
\begin{itemize}
\item Simulation domain:
\begin{center}
\begin{tabular}{rllll}
  1D:& $\quad x \in [-0.5; +0.5],$ & $\quad y = 0,$ &$\quad z = 0,$ &
  $\quad d=1$ \\
  diagonal: & $\quad x \in [-0.5; +0.5],$ & 
    $\quad y \in [-0.5; +0.5],$ & $\quad z=0,$ & $\quad d'=\sqrt{2}$
\end{tabular}
\end{center}
\item Grid: $x^i = -0.5 + (n-\frac{1}{2}) dx, \quad n=1\ldots 50\rho,
  \quad dx=1/(50\rho), \quad \rho \in \mathbb{Z}$
\item Time step: $dt = dx/4 = 0.005 / \rho$
\end{itemize}

As in the gauge wave case, the 1D evolution is carried out for
$T=1000$ crossing times, i.e.~$2\times10^5\rho$ time steps (or until
the code crashes), with output every 10 crossing times.  The 2D
diagonal runs are carried out for $T=100$, with output every crossing
time.  For the trivial directions ($y$ and $z$ for the wave
propagating along the $x$ axis and $z$ for the wave propagating along
the diagonal) we use the minimum number of grid points in the $y,z$
directions that allow for non-trivial numerical second derivatives.
For standard second order finite differencing this implies that we use
3 points in the appropriate directions. We run using $\rho=1,2,4$.


\begin{figure}
  \includegraphics[width=25pc]{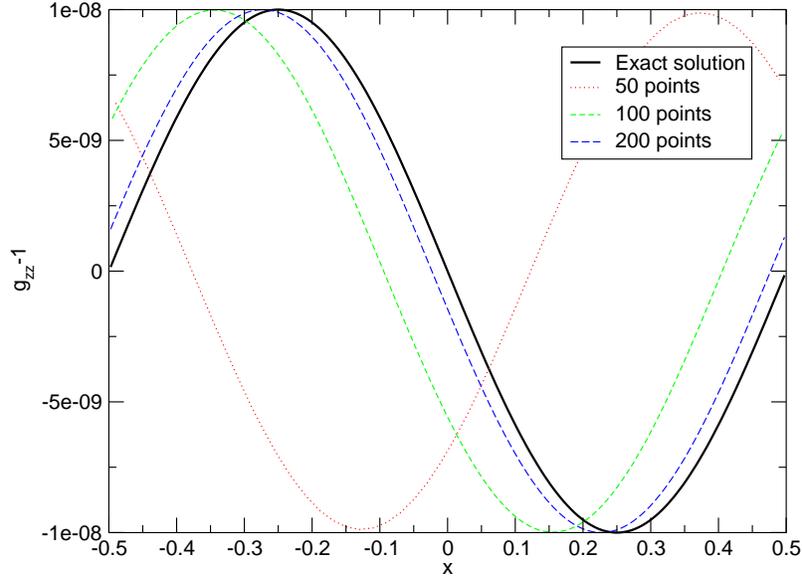}
  \caption{A 1D linear wave shown at different
    resolutions. Although the run lasted 1000 crossing times,
    the output is shown after 500 crossing times in order to
    indicate the trend of how resolution effects phase accuracy. 
    The numerical dissipation is low but the cumulative phase error
    is high at the coarser resolutions. It is clear that the phase error
    converges away. The results are from the code ABIGEL which
    implements a fully harmonic formulation~\cite{Szilagyi02a}.}
  \label{fig:LinearWave_gyy}
\end{figure}

The output quantities are similar to those for the gauge wave: the
$L_{\infty}$ and $L_2$-norms, the maxima and minima, and profiles along the $x$-axis
through the center of the grid of $g_{yy}$, $g_{zz}$, 
$tr(K)$, the Hamiltonian and any other nontrivial constraints, and the
$L_{\infty}$-norm of the difference from the linear exact solution for
$g_{zz}$. Figure \ref{fig:LinearWave_gyy} illustrates the profiles of
$g_{zz}-1$ obtained using a code based upon harmonic coordinates.

\subsection{Polarized Gowdy wave testbed}

All of the tests described so far considered initial data which were
perturbations of a flat background. Here we use a genuinely curved exact
solution -- a polarized Gowdy spacetime -- to test codes in a strong
field context.  The polarized Gowdy $T^3$ spacetimes are solutions of
the vacuum Einstein equations on the 3-torus, and describe an
expanding universe containing plane polarized gravitational
waves~\cite{Gowdy71}.  Gowdy spacetimes have previously been used for
testing numerical relativity codes by a number of
authors~\cite{MVP97,New98,Garfinkle02}. They have also extensively
been studied in mathematical cosmology; see e.g.~\cite{Ringstrom03}
for latest results.  An extensive analytical and numerical study of
Gowdy spacetimes has been carried out by Berger~\cite{Berger02}.

The polarized Gowdy metric is usually written as
\begin{equation}
  {ds}^2 = t^{-1/2} e^{\lambda/2} (-{dt}^2 + {dz}^2) + t\,{dw}^2,
  \label{metric-gowdy}
\end{equation}
where
\begin{equation}
  {dw}^2=e^P {dx}^2 + e^{-P}{dy}^2.
  \label{dw-pol-gowdy}
\end{equation}
Here the time coordinate $t$ is chosen such that time increases as the
universe expands. For code testing, it is quite interesting to compare
collapsing and expanding situations. We will thus carry out our tests
in {\em both} time directions.  The quantities $\lambda$ and $P$ are
functions of $z$ and $t$ only and are periodic in $z$.  The metric is
singular at $t=0$ which corresponds to the cosmological singularity.

With the metric (\ref{metric-gowdy}), the Einstein evolution equations can be
reduced to a single linear equation for $P$ \cite{Gowdy71}:
\begin{equation}
  P_{,tt} + t^{-1} \, P_{,t} - P_{,zz} = 0 \label{p_evolution}.
\end{equation}
The constraint equations become
\begin{equation}
  \lambda_{,t} = t\, (P_{,t}^2 + P_{,z}^2)\label{lambda-hamiltonian-constraint}
\end{equation}
and
\begin{equation}
  \lambda_{,z} = 2 \, t \, P_{,z} \, P_{,t} \label{lambda-momentum-constraint},
\end{equation}
and correspond to the Hamiltonian (\ref{lambda-hamiltonian-constraint}) and 
momentum (\ref{lambda-momentum-constraint}) constraints.
The general solution to eq.~(\ref {p_evolution}) is a sum of
terms of the form $\alpha \log t + \beta$, where $\alpha$ and $\beta$ are
real constants, and terms of the form $Z_0(2\pi nt)\cos(2\pi nz)$ and $Z_0(2\pi
nt)\sin(2\pi nz)$, where $n$ is an integer (assuming periodicity of 1
in $z$) and $Z_0$ is a linear combination of the Bessel functions $J_0$
and $Y_0$.
We follow \cite{New98} in the choice of the particular solution and set
\begin{equation}
  P = J_0(2\pi t)\cos(2\pi z), \label{def_P}
\end{equation}
which yields
\begin{equation}
  g_{xx}=te^P,\ g_{yy}=te^{-P},\ g_{zz}=t^{-1/2}e^{\lambda/2},
  \label{g_ij-polarizedgowdy}
\end{equation}
\begin{eqnarray}
  K_{xx}&=&- \frac{1}{2} t^{1/4}
  e^{-\lambda/4}e^P(1+t P_{,t}),
  \nonumber \\
  K_{yy}&=&- \frac{1}{2} t^{1/4}
  e^{-\lambda/4}e^{-P}(1 - t P_{,t}),
  \label{K_ij-polarizedgowdy} \\
  K_{zz}&=& \frac{1}{4} t^{-1/4}
  e^{\lambda/4}(t^{-1}-\lambda_{,t}).
  \nonumber
\end{eqnarray}
The shift vanishes, and the lapse is given as
\begin{equation}
  \alpha=\sqrt{g_{zz}} = t^{-1/4}e^{\lambda/4}.
  \label{alpha-polarizedgowdy}
\end{equation}
Using our choice for $P$ (\ref{def_P}), the constraint
eqs.~(\ref{lambda-hamiltonian-constraint},\ref{lambda-momentum-constraint}) 
yield an expression for $\lambda$:
\begin{equation}
  \label{eq:lambda}
  \begin{array}[c]{r c l}
  \lambda &=& -2\pi t J_{0}(2\pi t) J_{1}(2\pi t) \cos^{2}(2\pi z)
   + 2\pi^{2}t^{2} \bigl[J_{0}^{2}(2\pi t) + J_{1}^{2}(2\pi t)\bigr]
   \\
   & & \mbox{} - {\frac{1}{2}}
   \bigl\{ (2\pi )^{2}\bigl[J_{0}^{2}(2\pi )
   +J_{1}^{2}(2\pi )\bigr]-
   2\pi J_{0}(2\pi ) J_{1}(2\pi )\bigr\}.
  \end{array}
\end{equation}
Note that $A = \int_z t P_{,t} \, dz$ is a constant of motion and can
be used to monitor the accuracy of a code. In our case $A$ is set to
zero by the choice of initial data.

\begin{figure}
\begin{center}
  \includegraphics[width=20pc]{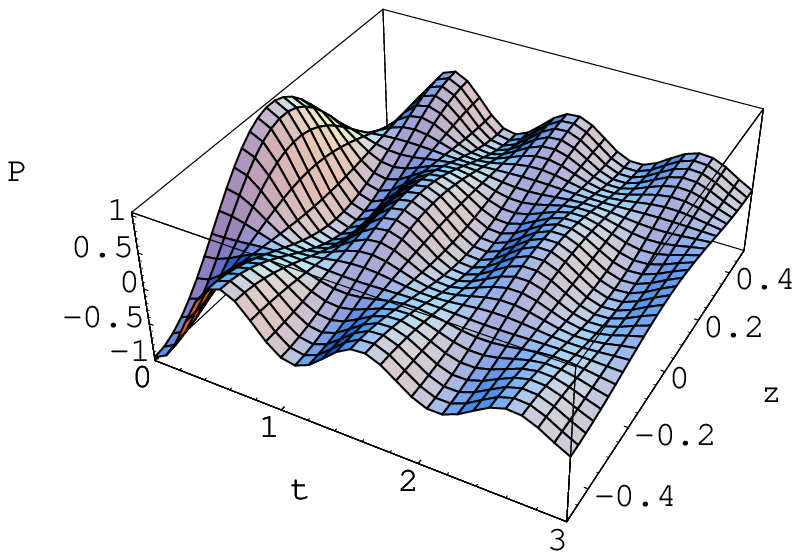}
%
  \includegraphics[width=20pc]{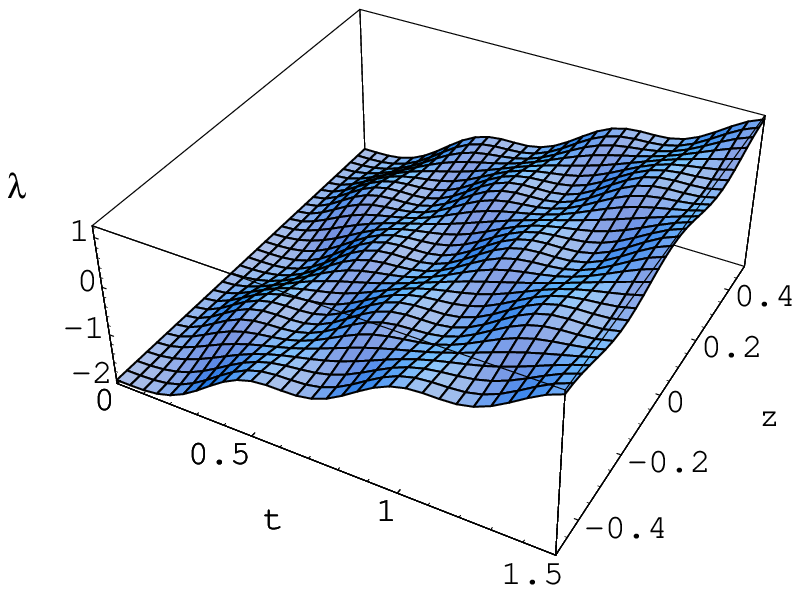}
  \caption{The quantities $P$ and $\lambda$ appearing in the
  Gowdy metric (\ref{metric-gowdy}) are plotted versus $z$ and $t$.}
  \label{fig:GowdyPLambda}
\end{center}
\end{figure}
\begin{figure}
\begin{center}
  \includegraphics[width=30pc]{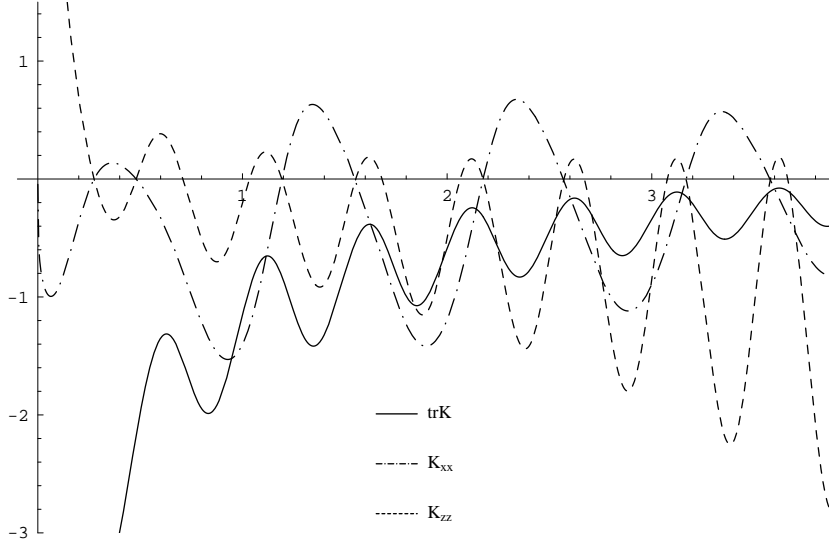}
  \caption{The extrinsic curvature components $K_{xx}$ and $K_{zz}$,
  as given in Eqs. (\ref{K_ij-polarizedgowdy}), 
  and $trK$ at the coordinate origin are plotted versus time $t$.}
  \label{fig:GowdyK}
\end{center}
\end{figure}

Figures showing $P$, $\lambda$ and extrinsic curvature components,
constructed from the analytic formulas, are given in
Figures~\ref{fig:GowdyPLambda} and \ref{fig:GowdyK}. While $P$ slowly
decays to zero, $\lambda$ shows a secular linear growth due to the
cosmological expansion, and both $P$ and $\lambda$ exhibit
gravitational wave oscillations. Note that although the individual
extrinsic curvature components do {\em not} exhibit a fixed sign,
$trK$ is negative and decays in absolute value, consistent with the
cosmological expansion.  The linear growth of $\lambda$ leads to
exponential growth in the metric component $g_{zz}$. This makes an
evolution with standard $3+1$ ADM variables much harder than evolving
the Gowdy quantity $P$.

The (coordinate) velocity of light is constant in the coordinates chosen
in eq.~(\ref{metric-gowdy}), and with a fixed spatial discretization size
$\Delta z$ the Courant
condition is consistent with a fixed timestep $\Delta t$. This makes
it convenient to choose the gauge (\ref{metric-gowdy}) for evolving in the
{\em expanding} direction. We will see below however, that this leads to
exponential growth in the metric component $g_{zz}$.
For the collapsing direction, this would lead to a singularity at $t=0$,
so we will evolve this case with a different slicing as discussed below.

For the forward (expanding) evolution, we set initial data from the
exact solution at $t=1$, which yields initial data of order unity, and
evolve with any lapse condition which is equivalent in the continuum
limit to the exact lapse given by eq.~(\ref{alpha-polarizedgowdy}).
Due to the exponential growth in the metric variables, such evolutions
may crash rather soon but will test the accuracy of a code in a rather
harsh situation. In order to evolve in the backward time direction, we
choose harmonic time slicing, as has previously been done by Garfinkle
\cite{Garfinkle02}. Since harmonic slicing is marginally singularity
avoiding \cite{Bona97a,Alcubierre02b}, such evolutions should only
asymptotically reach the singularity at $t=-\infty$.

It turns out that it is actually quite simple to write down an exact solution
for harmonic slicing, which greatly simplifies the task of choosing
appropriate gauge source functions for various formalisms.
Starting with the Gowdy metric, as given by eq.~(\ref{metric-gowdy}),
we look for a coordinate transformation $(t,x^i) \rightarrow (\tau, x^i)$,
with $t = F(\tau)$. In the new coordinates, the lapse becomes
$\hat \alpha = F(\tau)^{-1/4} \partial_\tau F(\tau) e^{\lambda/4}$.
The harmonicity condition $\Box \, t = 0$ then implies
\begin{equation}
e^{-\lambda/2} \left[ F \partial_{\tau\tau} F - 
\partial_\tau F^2 \right] = \sqrt{ F}  \partial_\tau F^3
\end{equation}
with the solution $F(\tau) = k e^{c \tau}$,
where $c$ and $k$ are free constants.
The lapse in this new gauge is
\begin{equation}
\hat \alpha(\tau) = c k^{3/4} e^{3 c \tau / 4 + \lambda(F(\tau),z) / 4}.
\end{equation}
In order to start the collapse slowly, and to simplify initial data, we
choose the constants $c, k$ in such a way that  $\hat \alpha = 1 $
at the initial time $t=t_0$.
Picking a value $t_0$ for which $J_0(2 \pi t_0) = 0$, eq.~(\ref{eq:lambda})
implies that $\hat \alpha$ is independent of $z$.
Using
$$
\tau_0 = \frac{1}{c} \ln \left( \frac{t_0}{k} \right), \quad
\lambda(k e^{c \tau_0}, z) = \lambda_0
$$
we obtain
\begin{equation}
\hat \alpha_0 = c\; t_0^{3/4}\; e^{\lambda_0 / 4}.
\end{equation}
Given our requirement $\hat \alpha_0 = 1$, and choosing 
$t_0 = \tau_0$, i.e.\ $F(\tau_0) = \tau_0$, we get
\begin{equation}
c = t_0^{-3/4}\; e^{-\lambda_0 / 4}, \quad
k = t_0 e^{-c t_0}.
\end{equation}
We will choose a particular value of $t_0$ such that the initial slice
is far from the cosmological singularity, but not so far that we have
to deal with extremely large numbers.  We pick the $20^{\text{th}}$
zero of the Bessel function $J_0(2 \pi t_0)$, which yields $t_0 \sim
9.8753205829098$, corresponding to
$$
c \sim 0.0021195119214617, \quad k \sim 9.6707698127638.
$$
The values of the metric components found from (\ref{metric-gowdy})
at $t = t_0$ are then $g_{xx} = g_{yy} = t_0$, $g_{zz} \sim 2.283
\times 10^3$.  This choice challenges a numerical code to accurately
track a small effect (the dynamics in $g_{xx}$, $g_{yy}$) together
with a larger effect (dynamics in $g_{zz}$). Other choices are of
course possible, and certainly worth exploring. For the purpose of a
standard testbed, which should provide tests which are able to
discriminate well between different formulations, the current choice
seems appropriate. 

The geometry of the grid is chosen analogous to the 1D gauge wave test:
\begin{itemize}
\item Simulation domain: $z \in [-0.5; +0.5]$,  $x = y = 0$
\item Grid: $z^i = -0.5 + (n-\frac{1}{2}) dz, \quad n=1\ldots 50\rho,
  \quad dz = 1/(50\rho), \quad \rho \in \mathbb{Z}$
\item Time step: $dt = dz/4 = 0.005 / \rho$
\item Run time: $T=1000$, i.e.~1000 crossing times or until code crash.
\end{itemize}

We output the $L_{\infty}$ and $L_2$-norms, the maxima and minima, and profiles
along the $z$-axis through the center of the grid of $g_{zz}$, $\alpha$,
$tr(K)$, the Hamiltonian constraint and all other nontrivial constraints
of the formulation, and some typical evolution variables, depending on
the evolution system chosen. We output norms every crossing time, and
profiles either every 10 crossing times or once per crossing time for
some initial time, depending on the behavior of the solution. 
We also calculate the $L_{\infty}$-norms of the difference from the exact
solution for $g_{xx}$ and $g_{zz}$ for the expanding direction.  

\begin{figure}
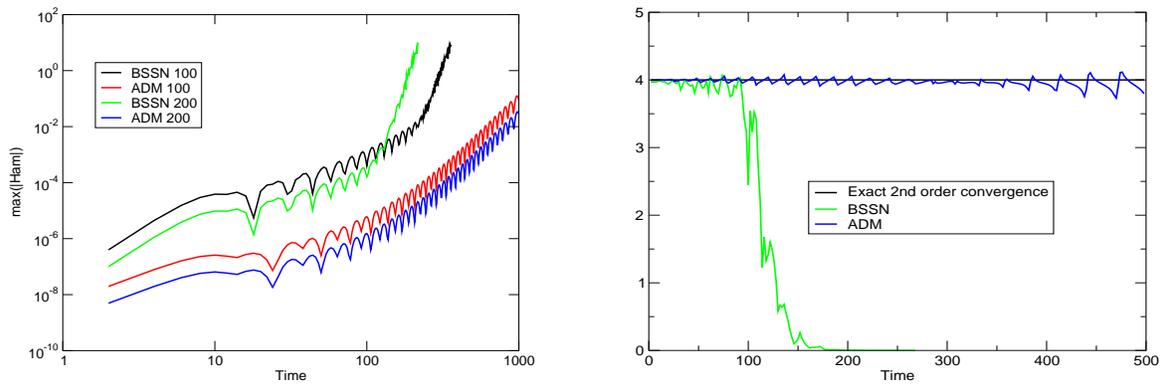

  \includegraphics[width=7.0cm,height=5cm]{figures/Gowdyconstraintgrowth.eps}
\hskip 1.2cm
  \includegraphics[width=7.0cm,height=5cm]{figures/ham_100_200_convergence.eps}
  \caption{Comparison of the $L_\infty$ norm of the Hamiltonian
constraint for ADM vs.\ BSSN. For the purpose of presentation the time 
coordinate has been adjusted to coincide with the number of crossing times. The
left figure shows the growth of the $L_\infty$ norm of the Hamiltonian
constraint on a double  logarithmic scale, the right figure shows the ratios of
the $L_\infty$ norm of the Hamiltonian  constraint for resolutions of 100 and
200 points. A value of $4$ would mark exact second order convergence.}
  \label{fig:GowdyCollapse_ADMvsBSSN}
\end{figure}

As a sample result we present a comparison of an ADM and a BSSN code
for the collapsing direction in
Figure~\ref{fig:GowdyCollapse_ADMvsBSSN}. While the ADM code shows
roughly second order convergence for 1000 crossing times (we show the
first 500 for better comparison with the BSSN results), only the
lowest resolution BSSN run lasts for 1000 crossing times with the
higher resolution runs crashing significantly earlier. The loss of
convergence is clear in Figure~\ref{fig:GowdyCollapse_ADMvsBSSN}. The
poor performance of the BSSN code seems to be rooted in its mixing of
components. The comparatively good performance of the ADM code
supports the usefulness of this test. Alternative choices of initial
data can be made to yield tests with different characteristics, but
will not be included in this round of tests.

\section{Discussion}

We have presented a program to develop a suite of standardized testbeds for
numerical relativity, and a first round of tests. All tests are specified in
great detail which extends to numerical methods, grid setup and choice of
output quantities to facilitate comparisons. The tests are based on vacuum
solutions and periodic boundaries. Even in this simple setup, the design of
tests is a highly nontrivial task and several subtle issues require careful
consideration: The effects of gravitational collapse introduce considerable
subtleties in tests of general relativity with periodic boundary conditions and
have been discussed in some detail. Comparing runs for spacetimes which possess
symmetries in different setups where the symmetry is manifest, disguised by a
coordinate transformation, or disguised by adding random noise can help to
understand the problem of what one can learn from simple one-dimensional tests,
and help test different aspects of a code. In particular, this has been found
useful in separating problems connected to ill-posedness from other sources of
instability or inaccuracy.

Several of the tests presented here have been used previously in one form or
another, but we  have tried to improve their specifications in order to
increase their practical value. We have modified the robust stability test
based on random noise as presented in \cite{Szilagyi00a} to reduce
computational resources when comparing different resolutions and added such a
comparison as an integral part of the test. Our setup of the collapsing
polarized Gowdy wave test combines a particularly simple choice of  initial
data with a simple form of the exact solution. 

The art of interpreting testbed results requires mastery of the art of
interpreting spacetimes. The latter has to be applied both to the continuum
limits and to the discretized approximations in order to understand results. A
simple example is provided by the gauge wave test, where individual runs may
exhibit collapse or expansion as a result of a physical instability of the
exact solution. Clearly, a valid code still has to show convergence to this
unstable exact solution. We strongly emphasize the importance of comparing
results for different resolutions. In particular, convergence tests not only
exhibit plain coding errors or numerical instabilities, but it is important to
obtain convergence information for all simulations individually, for the whole
length of a run. This is illustrated by our comparison of an ADM and a BSSN
code for the collapsing Gowdy test. Also, we emphasize that it is not
sufficient to monitor constraints to analyze instabilities,   but further
quantities need to be analyzed to render possible scientifically valuable
conclusions.

We have carried out sufficient experimentation with these tests to ensure that
they can be implemented with reasonable computational resources and that they
can effectively discriminate between the performance of different codes. A
separate paper presenting and interpreting test results for codes of all groups
that wish to participate will be prepared at a later date. At present, we
invite all numerical relativity groups to submit results and join as co-authors
in this  next paper.

Information on submitting results can be found at the web site
www.ApplesWithApples.org. Instructions can also be found there for accessing
the results submitted by the various participating groups. We also encourage
groups to submit results from tests that go beyond the ones proposed here and
that reveal further insight into code performance. This would be particularly
helpful in the design of future tests.  Also, information concerning
forthcoming workshops, and contact information for the participating groups,
are posted on the website.

The tests presented here are not intended to be an exhaustive or even
minimal list of tests that should be applied to a particular formulation or
code. However, they are sufficiently simple and general to allow all groups
to compare results with reasonable computational effort. They provide a way
of rapidly checking the utility of a code or formulation in situations
where detailed theoretical analysis is not possible. The tests also allow
isolation of problems of different origin, such as the mathematical  
formulation, the choice of gauge or the inaccuracy of the numerical method.
They do this in a simple situation where cross-comparison with other codes
can suggest remedies. 

We are proposing here the first step toward establishing a community wide
resource which will allow all groups to profit from each other's successes
and failures. Broad participation is essential to the success of this goal.
Future workshops, along the lines of the first Mexico workshop, are being
planned. The key challenge for the next round of tests will be to
include the significantly more complex problem of boundaries.

\begin{acknowledgments}

We have benefited from discussions with our colleagues T. Baumgarte,
H. Friedrich, R.~A.  Issacson, L. Lindblom and B. Schmidt. We are especially
indebted to A. Rendall and H. Ringstr\"{o}m for helping identify some
of the traps in designing useful testbeds with toroidal topology. Many
of the groups participating in these tests have utilized the Cactus
infrastructure~\cite{Goodale02a}. The work was supported by NSF grant
PHY 9988663 to the University of Pittsburgh and PHY-0071020 to
the University of Maryland, the Collaborative
Research Centre (SFB) 382 of the DFG at the University of T\"ubingen. 
H. Shinkai is supported by the special
postdoctoral researcher program at RIKEN and partially by the
Grant-in-Aid for Scientific Research Fund of the Japan Society for the
Promotion of Science, No.~14740179. D.~Shoemaker acknowledges the
support of the Center for Gravitational Wave Physics funded by NSF PHY-0114375
and grant PHY-9800973. M.~Alcubierre acknowledges partial
support from CONACyT-Mexico through the repatriation program and from
DGAPA-UNAM through grants IN112401 and IN122002. R.~Takahashi thanks
A. Khokhlov and I. Novikov for encouragement and hospitality,
S. Husa thanks the university of the Balearic islands for hospitality.

\end{acknowledgments}



\bibliographystyle{apsrev}
\bibliography{mexico1}

\end{document}